\documentstyle[prl,aps]{revtex}

\begin{document}

\newcommand{\be}{\begin{equation}}
\newcommand{\ee}{\end{equation}}
\newcommand{\bear}{\begin{eqnarray}}
\newcommand{\eear}{\end{eqnarray}} \newcommand{\ba}{\begin{array}}
\newcommand{\ea}{\end{array}}
\newcommand{\lae}{\begin{array}{c}\,\sim\vspace{-1.7em}\\<
\end{array}}
\newcommand{\gae}{\begin{array}{c}\,\sim\vspace{-1.7em}\\>
\end{array}}

\newcommand{\CA}{{\cal A}} \newcommand{\CQ}{{\cal Q}}
\newcommand{\CU}{{\cal U}} \newcommand{\CD}{{\cal D}}
\newcommand{\CL}{{\cal L}} \newcommand{\CE}{{\cal E}}
\newcommand{\CN}{{\cal N}} \newcommand{\hg}{\hat{g}}
\newcommand{\CH}{{\cal H}} \newcommand{\notD}{\not\!\! D}

\def\vbr{$\vphantom{\sqrt{F_e^i}}$}

\twocolumn[\hsize\textwidth\columnwidth\hsize\csname
@twocolumnfalse\endcsname

\title{Proton Stability in Six Dimensions}

\author{Thomas Appelquist,
Bogdan A.~Dobrescu, Eduardo Pont\'{o}n, Ho-Ung Yee}

\address{Department of Physics, Yale University, New
Haven, CT 06511, USA
}

\date{July 5, 2001} \maketitle

  \begin{abstract}
{\small We show that Lorentz and gauge invariance
explain the long proton lifetime within the
standard model in six dimensions.
The baryon-number violating operators have mass
dimension 15 or higher. Upon TeV-scale compactification
of the two universal extra dimensions on a square $T^2/Z_2$
orbifold, a discrete subgroup of the 6-dimensional Lorentz
group continues to forbid dangerous operators. }
\end{abstract}

\vspace*{-5.2cm}
\noindent hep-ph/0107056  \hfill YCTP-P5-01  \\
\vspace*{4.8cm}
]
\setcounter{footnote}{0}
\setcounter{page}{1}

The stability of the proton is one of the most intriguing problems
in physics. The current limit on the proton lifetime is $1.6\times
10^{33}$ years if the dominant decay mode is $p \rightarrow e^+
\pi^0$, and $1.6\times 10^{25}$ years independent of the decay
mode \cite{Groom:2000in}. The conserved quantity associated with
proton stability, the baryon number $B$, is the charge
corresponding to a global $U(1)_B$ symmetry, or one of its
discrete subgroups. It is important to explain why $B$ is
conserved to such a high degree of accuracy.  In this letter we
demonstrate that the spatial symmetry of two universal extra
dimensions, accessible to all the standard model fields,
naturally constrains the proton lifetime to be
greater than the experimental limits.

The minimal standard model, including operators of dimension four
and lower, automatically conserves $B$. This theory, however, is
valid only up to some scale $\Lambda_{\rm SM}$ where new physics
becomes manifest. The effect of physics at scales above
$\Lambda_{\rm SM}$ is described most generally by higher dimension
operators, suppressed by the appropriate power of $\Lambda_{\rm
SM}$. Among these are $B$ violating operators of dimension six:
$(\overline{q}^c_L q_L)(\overline{q}^c_L l_L)$,
$(\overline{d}^c_R u_R)(\overline{q}^c_L l_L)$,
$(\overline{q}^c_L q_L)(\overline{u}^c_R e_R)$,
$(\overline{u}^c_R d_R)(\overline{e}^c_R u_R)$, where only one
Lorentz invariant form is displayed for each four-fermion term;
the weak-eigenstate quarks, $q_L$, $u_R$, $d_R$, and leptons,
$l_L$, $e_R$, may belong to any generation. The experimental
limits on the proton lifetime impose a bound on the coefficients
of these operators: $C_{\rm SM}/\Lambda_{\rm SM}^2 \lae 10^{-24}
{\rm TeV}^{-2}$. If $\Lambda_{\rm SM} \gae 10^{16}$ GeV, then the
coefficients $C_{\rm SM}$ can safely be as large as order one at
that scale. However, the naturalness of the Higgs sector in the
standard model restricts $\Lambda_{\rm SM}$ to lie in the TeV
range. The $C_{\rm SM}$'s must therefore be extremely small ---
below $10^{-23}$ or so.

Various explanations for the smallness of $B$ violation have been
proposed. In the minimal supersymmetric standard model, a discrete
symmetry is needed to banish $B$-violating
terms of dimension three and four from the superpotential.
The discrete symmetry may be a remnant of a
$U(1)$ gauge group \cite{Weinberg:1982wj}, or else one could
hope that it is preserved by quantum gravity.
In the case of large extra dimensions,
the smallness of $C_{\rm SM}$
has been suggested to arise from
the localization of the quarks and leptons at different points
inside a thick brane \cite{Arkani-Hamed:2000dc}.

Here we concentrate on the chiral 6-dimensional standard model, in
which all standard model fields propagate in two extra
dimensions. The current bound on the compactification scale of
two universal extra dimensions is $1/R \gae 500$ GeV
\cite{Appelquist:2001nn}, suggesting a rich phenomenology
\cite{Agashe:2001ra}. The standard model in two universal extra
dimensions is especially appealing because of the properties of
the Lorentz group in six dimensions, as well as the constraints
imposed by anomaly cancellation.
The 6-dimensional standard model is an
effective theory valid up to a scale $M_s \approx 5/R$, above which
the 6-dimensional $SU(3)_C\times SU(2)_W \times U(1)_Y$ gauge
interactions become non-perturbative.

In six dimensions, the $SO(1,5)$ Lorentz symmetry has two
irreducible spin-1/2 representations generated by
$\Sigma^{\alpha\beta}/2$, where $\Sigma^{\alpha\beta} \equiv
i[\Gamma^\alpha, \Gamma^\beta]/2$ in terms of the anticommuting
$8\times 8$ matrices $\Gamma^\alpha$, $\alpha = 0,1, ...,5$. The
product  $\Gamma^0\Gamma^1...\Gamma^5$ has eigenvalues $\pm 1$
defining the two 6-dimensional chiralities. The irreducible gauge
[$SU(3)_C\times SU(2)_W \times U(1)_Y$] and gravitational
anomalies cancel for only two chirality assignments, up to an
overall sign \cite{Dobrescu:2001ae,Arkani-Hamed:2000hv}:
\be
\CQ_+ \; , \; \CU_- \; , \; \CD_- \;  , \;
\left\{
\begin{array}{ccc} \CL_+ \; , & \CE_- \; , & \CN_- \; , \\
& {\rm or} & \\ [0.7mm] \CL_- \;  , & \CE_+ \; , & \CN_+ \; ,
\end{array} \right.
\label{assign}
\ee where $\CQ_+, \CU_-, \CD_-$ and $\CL_\pm,
\CE_\mp, \CN_\mp$ are the 6-dimensional quarks and leptons,
respectively, and a generational index is implicit. We assume
that only one of these two assignments applies to all
fermion generations, since then the $SU(2)_W$
global anomaly demands 3 mod 3 generations \cite{Dobrescu:2001ae}.
Each fermion generation includes a gauge singlet field,
$\CN_\mp$, such that the gravitational anomaly cancels. The
6-dimensional chiral quark and lepton fields have four
components, and if two dimensions are appropriately compactified
on an orbifold, then their zero-modes may be identified with the left-
and right-handed standard model fermions $q_L, u_R, d_R, l_L,
e_R$, as well as three right-handed neutrinos $\nu_R$.
We first describe the constraints on $B$ violation from 6-dimensional
Lorentz invariance, and then show that the constraints are
equally tight even with the lesser symmetry remaining after orbifold
compactification.

We begin by observing that $SO(1,5)$ has an $SO(1,3) \times
U(1)_{45}$ subgroup: the combination of the 4-dimensional
Lorentz symmetry associated with the $x^\mu$ coordinates,
$\mu = 0,1,2,3$, and the symmetry under rotations of the
$x^4$, $x^5$ coordinates. Any chiral 6-dimensional fermion,
$\Psi_\pm$, may be decomposed under this subgroup as
\be \Psi_\pm = {\Psi_\pm}_L + {\Psi_\pm}_R, \ee
\begin{figure}[t]
\begin{center}
\renewcommand{\arraystretch}{1.4}
\begin{tabular}{|c|c|c|}\hline
\ Fermion \ & \ $U(1)_{45}$ charge \ & \ zero-mode \ \vbr\vbr \\ \hline\hline
${\CQ_+}_L$   & $-1/2$ & \ $q_L = (u_L, d_L)$ \vbr\vbr\\ \hline
\ ${\CU_-}_R$, ${\CD_-}_R$  \  & $-1/2$ & $u_R$, $d_R$  \vbr\vbr\\ \hline
\ ${\CQ_+}_R$, ${\CU_-}_L$, ${\CD_-}_L$   & $+1/2$ & --- \vbr\vbr\\ \hline
${\CL_\pm}_L$ & $\mp 1/2$ & \ $l_L=(\nu_L, e_L)$  \vbr\vbr\\ \hline
${\CE_\mp}_R$, ${\CN_\mp}_R$ & $\mp 1/2$ & $e_R$, $\nu_R$  \vbr\vbr\\ \hline
${\CL_\pm}_R$, ${\CE_\mp}_L$, ${\CN_\mp}_L$ & $\pm 1/2$ &  ---
\vbr\vbr \\ \hline
\end{tabular}
\bigskip \\
\parbox[t]{8.2cm}{\small TABLE I. Transformation properties of the
quark and lepton fields under $SO(1,3)\times U(1)_{45}$,
and the corresponding zero-modes under orbifold compactification.}
\end{center}
\end{figure}
\noindent
where $L$ and $R$ are $SO(1,3)$ chiralities, defined by the
projection operators
\be {P_\pm}_L
= {P_\mp}_R = \frac{1}{2}\left( 1 \mp \Sigma^{45} \right) ~.
\ee
Since $\Sigma^{\alpha\beta}/2$ are the generators
of the spin-1/2 representations of $SO(1,5)$,
the $SO(1,3)$-chiral fermions have $U(1)_{45}$ charges given
by the eigenvalues of $\Sigma^{45}/2$:
$\mp 1/2$ for ${\Psi_\pm}_L$, and $\pm 1/2$ for ${\Psi_\pm}_R$. In
Table I we list the charges of the quarks and leptons. The
covariant derivative, $D_\alpha$, transforms as a vector under
$SO(1,5)$, so that $D_4 + i D_5$ has $U(1)_{45}$  charge $-1$.

To construct the $B$-violating operators in six dimensions, we
first note that since all quark fields carry $B=+1/3$, $SU(3)_C$
gauge invariance demands that for any such operator the number of
quark fields (minus the number of charge-conjugated fields) is a
multiple of three. Only operators constructed of fields that have
zero modes can induce at tree-level the $B$-violating processes
searched for in experiments so far. From Table I we see that all
quark fields containing zero modes have $U(1)_{45}$ charge $-1/2$, so
that an operator with $|\Delta B| \ge 1$ has a $U(1)_{45}$ charge
given by $(-3/2)\Delta B$ plus the sum of the charges of all lepton
fields included in that operator. Hence, {\it all $B$-violating
nucleon-decay
operators allowed by Lorentz invariance in six dimensions must
involve at least three quarks and three leptons.}

Using this key fact, it is straightforward to find the lowest
dimension operators, invariant under $SO(1,5)$ and
standard-model gauge transformations, that (after compactification)
can induce $B$-violating nucleon decays. We
consider first the $\CL_+$ chirality assignment. The
operators then first appear at mass-dimension 16 in the
6-dimensional theory, each involving six fermions and one
covariant derivative $\notD = D_\beta\Gamma^\beta$:
\bear
&& \left( \overline{\CL}_{+} \CD_{-} \right)^2
\left( \overline{\CN}_{-} \notD \CD_{-} \right) ~,
\; \left( \overline{\CE}_{-} \notD \CD_{-} \right)
\left( \overline{\CN}_{-} \Gamma^\alpha \CD_{-} \right)^2  ~,
\nonumber \\
&& \left( \overline{\CL}_{+} \CD_{-} \right)
\left( \overline{\CN}_{-} \CQ_{+} \right)
\left( \overline{\CN}_{-} \notD \CD_{-} \right) ~,
\nonumber \\
&& \left( \overline{\CN}_{-} \notD \CU_{-} \right)
\left( \overline{\CN}_{-} \Gamma^\alpha \CD_{-} \right)^2 ~,
\; \left( \overline{\CN}_{-} \CQ_{+} \right)^2 \left(
\overline{\CN}_{-} \notD \CD_{-} \right) ~, \label{dim-16}
\eear
where we have exhibited only the Lorentz covariant bilinears with
the smallest number of $\Gamma$ matrices; permutations of $\notD$
and $\Gamma^\alpha$ are also allowed. The $SO(1,3)$ chiralities of
the zero modes imply that only the terms in (\ref{dim-16}) which
do not contain $\Gamma^{4,5}$ can induce nucleon decay. Each of
the above dimension-16 operators enters with a coefficient
proportional to $M_{s}^{-10}$ leading to a strong (and adequate)
suppression of $B$ violation after compactification to four
dimensions. We estimate their effects after first
considering dimension-17 operators, whose contributions
to proton decay turn out to dominate.

There is only one proton-decay operator of
dimension 17 (modulo Fierz transformations, and
the insertion of two or three $\Sigma^{\alpha\beta}$ matrices),
invariant under $SO(1,5)$ and standard model gauge transformations
(in particular $U(1)_Y$),
that does not involve the gauge singlets $\CN_-$:
\be {\cal O}_{17} = \frac{C_{17}}{M_s^{11}} \left(
\overline{\CL}_{+} \CD_{-} \right)^3 \tilde{\CH} ~,
\label{o17}
 \ee where
$\tilde{\CH}$ is the charge-conjugated Higgs doublet in six
dimensions. Dimension-17 operators involving $\CN_-$ have a
similar form, and will be discussed below. All other
$B$-violating nucleon-decay operators require more derivatives,
$\tilde{\CH}$ fields, or fermion bilinears.

Upon integrating over the compact dimensions, $x^4$, $x^5$,
restricting the fields to their zero modes, and replacing the
zero-mode Higgs doublet by its VEV ($v_h\approx 174$ GeV),
${\cal O}_{17}$ gives rise in the 4-dimensional theory to the operator
\be
\frac{v_h C_{17}}{A_{45}^{5/2} M_s^{11}} \left(
\overline{\nu}_L d_R \right)
\left( \overline{l}_L d_R \right)^2 ~,
\ee
where $A_{45}$ is the area spanned by $x^4$ and $x^5$.
For the $T^2/Z_2$ square
orbifold of radius $R$ constructed in \cite{Appelquist:2001nn},
$A_{45} = 2\pi^2 R^2$.
This operator is non-vanishing only if the generational indices
of the three $\overline{\CL}_+ \CD_-$ bilinears in ${\cal O}_{17}$
are not all identical. Hence, ${\cal O}_{17}$
induces proton decays into $e^-\pi^+\pi^+\nu \nu$
or $\mu^-\pi^+\pi^+\nu\nu$ with the two neutrinos belonging
to different generations. Final states
with more pions or a $K^+$ are kinematically suppressed. The proton
width is given by:
\be
\Gamma( p\rightarrow e^-\pi^+\pi^+\nu\nu ) \approx
\frac{v_h^2 C_{17}^2 }{A_{45}^5M_s^{22}}
\left(\frac{m_p}{3}\right)^{11} \Phi_5 F(\pi\pi) ~,
\label{width}
\ee
where $m_p$ is the proton mass.
If the process occurs predominantly via the decay of the $d$ valence quark
\cite{Jarlskog:1979uu}, whose constituent mass $m_p/3$ is used
to set the scale in Eq.~(\ref{width}), then the form factor
$F(\pi\pi)$, which describes the probability for the two pions to
be formed, is expected to be of order unity.
The kinematical phase-space factor,
$\Phi_5$, defined here as the dimensionless quantity
that relates the squared amplitude to the decay width,
is tiny for a five-body decay.
We estimate $\Phi_5 \lae (4\pi)^{-7} \times O(10^{-4})$, where the factors
of $4\pi$ account for angular integration and the additional suppression
is due to integration over the magnitudes of final state momenta.
The process $p\rightarrow \mu^-\pi^+\pi^+\nu\nu$
has an even smaller $\Phi_5$.
Thus, ${\cal O}_{17}$ yields a finite proton lifetime:
\be
\tau_p \approx \frac{10^{35} {\rm yr}}{C_{17}^2}
\left[ \frac{(4\pi)^{-7}10^{-4}}{\Phi_5 F(\pi\pi)} \right]
\left[\frac{1/R}{0.5 \; {\rm TeV}}\right]^{\! 12}
 \left[\frac{RM_s}{5}\right]^{\! 22} ,
\label{taup}
\ee
where the quantities in square brackets are of order one
or larger. There are no published experimental limits for
five-body proton decays.
For comparison, the searches for $p \rightarrow e^-\pi^+\pi^+$
\cite{Berger:1991fa} set a limit of $\tau_p > 3\times 10^{31}$ yr.
Thus, for $C_{17} \leq O(1)$, the proton lifetime is orders of
magnitude longer than the current experimental limits.

The ${\cal O}_{17}$ operator also induces $B$-violating neutron
decays, including $n\rightarrow e^-\pi^+\nu\nu$, $\mu^-\pi^+\nu\nu$.
The four-body decays proceed via
the fusion of two $d$ valence quarks, which is suppressed by
their wave function overlap inside the neutron
\cite{Jarlskog:1979uu}. This is compensated though by the larger
constituent mass, $\sim 2m_p/3$, of the $dd$ pair. Compared
with $\Gamma( p\rightarrow e^-\pi^+\pi^+\nu\nu )$ given in
Eq.~(\ref{width}), the neutron partial width in these modes is
enhanced by $\Phi_4/\Phi_5$, where $\Phi_4 \lae (4\pi)^{-5}
\times O(10^{-3})$ is the four-body phase-space factor. Thus, the
inverse width of the $B$-violating neutron decays induced by
${\cal O}_{17}$ is smaller than $\tau_p$ by three orders of
magnitude. The experimental limit on the partial mean life of
$n\rightarrow e^-\pi^+$ is $6.5 \times 10^{31}$ yr
\cite{Seidel:1988ut}. If a dedicated search for $n\rightarrow
e^-\pi^+\nu\nu$ were to yield a comparable limit, then only a
weak bound of $C_{17} \lae O(1)$ would be imposed, with
uncertainties especially due to the sensitive dependence on $M_s$
and $R$.

These are striking results. $B$-violating nucleon decays are
adequately suppressed even with the scale of $B$ violation in the
TeV range, providing only that the coefficient $C_{17}$ is not
larger than order unity. In fact, $C_{17}$ is likely to be
significantly smaller than unity because the operator ${\cal
O}_{17}$ is composed of  chirality flipping bilinears and
therefore can be expected to arise with strength proportional to
small Yukawa couplings.

We now return to the other operators.
For the dimension-16 operators of Eq.~(\ref{dim-16}),
the presence of the covariant derivative leads to an additional
factor of $m_{p}/3$ in the 4-dimensional amplitude in place of a
$A_{45}^{-1/2}v_h/M_{s}$ factor. Thus, there is a supression factor
of $10^3$ in the widths relative to those arising
from dimension-17 operators.
As we mentioned above,
all dimension-17 operators other than ${\cal O}_{17}$
involve at least one $\CN_-$ field. The tiny neutrino
masses suggest that the couplings of its zero mode, $\nu_R$, are
small. Thus, the coefficients of these operators may naturally
be substantially smaller than $C_{17}$. Nonetheless, some
could be relevant because they allow new decay modes:
$p \rightarrow \pi^+ \nu\nu\nu$, $n \rightarrow \pi^0 \nu\nu\nu$,
and other final states with mesons and three neutrinos.
[$n \rightarrow \nu\nu\nu$ requires a three-quark fusion
\cite{Jarlskog:1979uu} whose small probability is
not compensated by the gain in phase space.] The
experimental limits on such modes are of order $10^{32}$ yr
\cite{Groom:2000in}.
There exists an operator of this type with a single
$\CN_-$ field, $(\overline{\CL}_+ \CD_-)^2(\overline{\CN}_- \CQ_+)
\tilde{\CH}$, and a few with more $\CN_-$ fields.
The bounds on the coefficients of these
operators are as loose as those on $C_{17}$.

Consider next
the $\CL_-$ chirality assignment. It leads to a different set of
proton-decay operators, invariant under $SO(1,5)$ and
gauge transformations: the dominant ones are six-fermion
operators of dimension 15,
\be
\frac{1}{M_s^{9}}
\left[ C_{15}(\overline{\CE}_+^c \CU_-)
+ C_{15}^\prime(\overline{\CL}_-^c \CQ_+) \right]
\left( \overline{\CL}^c_{-} \Gamma^\alpha  \CU_{-} \right)^2~.
\label{o15}
\ee
Other six-fermion operators involve $\CN_+$ fields,
but again, their coefficients may be small.
The operators (\ref{o15}) lead to proton decay into three
anti-leptons (one or two are
positively charged) and a number of mesons. For the
dominant decay mode, $p\rightarrow e^+e^+\pi^-\bar{\nu}$,
we estimate
\be
\tau_p \approx \frac{10^{26} {\rm yr}}{C_{15}^2}
\left[ \frac{(4\pi)^{-5}10^{-3}}{\Phi_4 F(\pi)} \right]
\left[\frac{1/R}{0.5 \; {\rm TeV}}\right]^{\! 10}
 \left[\frac{RM_s}{5}\right]^{\! 18}\!\!\! ~,
\label{taupm}
\ee
where $F(\pi)$ is a form factor of order unity.
Despite the reduced phase space,
the process $p\rightarrow \mu^+e^+\pi^-\bar{\nu}$
is more constraining
due to better data on decays into muons \cite{Groom:2000in}:
$\tau_p/{\rm Br}(p\rightarrow \mu^+ X) > 10^{31}$ yr,
where Br$(p\rightarrow \mu^+ X)$ is the branching
fraction for inclusive decays. We derive a constraint,
$C_{15}
\lae 10^{-2} (R\times 0.5 \; {\rm TeV})^{-5}$,
which is rather loose given that
$C_{15}$ may naturally be as small as some Yukawa couplings.
Nonetheless, the prospects for observing $B$ violation look
better compared with the $\CL_+$
chirality assignment. Recall that
no known theoretical argument determines
which of the two chirality assignments of Eq.~(\ref{assign}) is
preferred.

We have so far considered proton decay arising
from operators that respect the $SO(1,5)$ symmetry. But since the
compactification of two dimensions breaks $SO(1,5)$, including its
$U(1)_{45}$ subgroup, we must next study whether the reduced
symmetry allows operators that could induce proton decay at an
unacceptable level. A simple and symmetric
choice for the compactification
of the $x^4$ and $x^5$ dimensions is a $T^2/Z_2$ orbifold of
equal radii \cite{Appelquist:2001nn}.
The ``torus'' $T^2$ is taken to be a square of size $2\pi R$
in the $x^4, x^5$ plane,
with periodic boundary conditions, $\varphi(x^4, x^5) =
\varphi(x^4 + 2\pi R, x^5) = \varphi(x^4, x^5 + 2\pi R)$ for any field
$\varphi$. The square has a $Z_4$ symmetry, being
invariant under $\pi/2$ rotations in the $x^4, x^5$ plane.
Now, since the generator of the $U(1)_{45}$ rotations
acting on fermions is $\Sigma_{45}/2$ [the spin-1/2 representation
is ``double-valued'' under $2\pi$ rotations], upon compactification on
a square, the $U(1)_{45}$ symmetry is broken down
to a $Z_8$ group whose elements are given by $U^n$ with
$U = \exp[i(\pi/2)\Sigma_{45}/2]$ and $n=0,1,..., 7$.

Compactification on $T^2$ would allow only
vectorlike zero-mode fermions in the 4-dimensional theory, so
it is necessary to introduce the observed $SO(1,3)$ chirality of the
standard model fermions by orbifolding the square.
Each field is taken to be either even or odd under
the $Z_2$
orbifold transformation, defined by a $\pi$ rotation in the
$x^4, x^5$ plane. The assigned $Z_2$ parity of the quarks and
%
\begin{figure}[t]
\begin{center}
\begin{picture}(-170,150)(100,-70)
\put(0,-60){\vector(0, 1){136}}
\put(-70,0){\vector(1, 0){160}}
\multiput(0,-0.3)(0, 2.56){16}{\line(1,0){40}}
\multiput(0,0.3)(0,-2.56){16}{\line(1,0){40}}
\multiput(0,-40)(2.5, 0){16}{\line(0,1){80}}
\thicklines
\put(-40,-40){\line(1, 0){80}}
\put(-40,-40){\line(0, 1){80}}
\put(40,40){\line(-1, 0){80}}
\put(40,40){\line(0, -1){80}}
\put(80,-12){$x^4$}
\put(-12,70){$x^5$}
\put(-24,-50){$- \pi R$}
\put(-65, -12){$- \pi R$}
\put(-16,46){$ \pi R$}
\put(45, -12){$\pi R$}
\put(0,40){\circle*{5}}
\put(40,0){\circle*{5}}
\put(0,0){\circle*{5}}
\put(40,40){\circle*{5}}
\end{picture}
\smallskip \\ [-2mm]
\parbox[t]{8.2cm}{\small FIG. I. The square $T^2/Z_2$ orbifold
of radius $R$: the fundamental region is shaded, and
the four fixed points under $Z_2$ are marked by $\bullet$.}
\end{center}
\end{figure}
\vspace*{-.44cm}
\noindent
leptons can be read from Table I: a field has a zero mode
if and only if it is even.
The fundamental region in the $x^4, x^5$ plane of the
square orbifold is shown in Fig.~I.

The important point is that the square $T^2/Z_2$ orbifold
does not break further the $Z_8$ symmetry of the 4-dimensional
effective theory.
This is because the $Z_2$ orbifold projection commutes with the
$Z_8$ transformation.
A fermion $\psi$ of charge $z$ under $Z_8$ transforms as follows:
\be
U \psi(x^4, x^5) = \sigma^k
e^{i z \pi/2} \psi(|x^5|, \sigma x^4) ~,
\ee
where $k=1$ ($k=2$) if $\psi$ is odd (even) under $Z_2$.
The transformation of $\psi(x^4, x^5)$ depends
on the sign of $x^5$, through
$\sigma = +1$ ($\sigma = -1$) for $x^5 < 0$ ($x^5 \ge 0$).
Note also that the presence of orbifold fixed points (see Fig.~1)
does not break the $Z_8$ symmetry.

Having established that the residual symmetry of the effective
theory is $SO(1,3) \times Z_8$, where $Z_8$ is the subgroup of
$U(1)_{45}$ defined above, it follows from $SU(3)_C$ gauge
invariance and from the charges given in Table I that all operators
constructed only of zero-mode fields satisfy the selection rule
$\frac{3}{2} \Delta B \pm \frac{1}{2} \Delta L = 0 \; {\rm mod}
\,4$. In particular, no proton-decay operators with less than 6
fermions are allowed and our previous operator analysis remains
valid (with the dominant $B$-violating effects given by
dimension-17 (-15) operators in the $\CL_+$ ($\CL_-$) chirality
assignment). Furthermore, it is also clear that in this model there
are no $n-\bar{n}$ oscillations (which require a $\Delta B
= 2$ operator) and no Majorana masses ($\Delta L = 2$).

The remarkable suppression of $B$ violation established here is
due to the $Z_8$ symmetry. But if, for example, $x^4$ and $x^5$
were compactified on a rectangular orbifold ($T^2/Z_2$ with
different radii), then the $Z_8$ would be reduced to a $Z_4$. It
would follow that dimension-10 operators such as
$(\overline{\CQ}_+^c \Gamma^\alpha\CQ_+) (\overline{\CQ}_+^c
\Gamma_\alpha\CL_+)$, which do not include terms with only
zero-modes, would induce proton decay via loops. [For the $\CL_-$
chirality assignment, the leading $B$-violating operators have
dimension 12, {\it e.g.} $(\overline{\CQ}_+^c \Gamma^\alpha\CQ_+)
(\overline{\CL}_-^c\Gamma_\alpha\CD_-)\tilde{\CH}$.] It is
therefore important to explore whether the $Z_8$ symmetry is indeed
left intact by the compactification. Here we offer only a few
comments regarding the choice of vacuum. It has been noted that the
square configuration is an extremum of the Casimir energy of bulk
fields, including gravity, while the rectangular one is not
\cite{Ponton:2001hq}. Although the lowest order computation
indicates that the square configuration is a saddle point of the
effective potential (the minimum corresponds to a ``rhombus" that
has a $Z_6$ symmetry, still forbidding $B$-violating operators with
less than three quarks and three leptons), it could be that a
complete computation will reveal this configuration to be a
minimum. Alternatively, it is possible to freeze the shape of the
torus to be square by a $Z_4$ orbifold identification, in which
case the $Z_8$ symmetry remains unbroken. Either of these
possibilities would leave intact our conclusions based on the
$T_2/Z_2$ square orbifold regarding $B$ violation.

We finally mention that some of the fields describing gravitational
fluctuations are also charged under the $Z_8$ (the components of
the metric along $x^4$ and $x^5$). However, we need not worry about
the effects of these fields for nucleon decay if the radius
stabilization mechanism makes them heavier than the proton or if
they are projected out by a $Z_4$ orbifold identification.

In conclusion, we have shown that the combination of
standard-model gauge invariance, 6-dimensional Lorentz
invariance, and compactification of the two extra dimensions on a
$T^{2}/Z_2$ orbifold of equal radii naturally suppresses proton
decay to acceptable levels, even with the scale of baryon number
violation in the TeV range.

\smallskip
We are grateful to Hsin-Chia Cheng
 and Erich Poppitz for illuminating conversations.
 This work was supported by DOE under contract DE-FG02-92ER-40704.

\end{document}